\documentclass[12pt,preprint]{aastex}

\slugcomment{Accepted for publication in ApJ, 2007}

\shorttitle{Phase-Resolved Infrared Spectroscopy of EF Eridani}
\shortauthors{Campbell et al.}

\begin{document}

\title{Cyclotron Modeling Phase-Resolved Infrared Spectroscopy of Polars I: EF Eridani}

\author{Ryan K. Campbell\altaffilmark{1,2} and
Thomas E. Harrison\altaffilmark{1}}
\affil{Astronomy Department, New Mexico State University,
 Las Cruces, NM 88003}

\author{Axel. D. Schwope}
\affil{Astrophysikalisches Institut Potsdam, An der Sternwarte 16, Potsdam,
  14482}

\author{Steve. B. Howell}
\affil{NOAO, 950 N. Cherry Ave., Tucson, AZ 85719}

\altaffiltext{1}{Visiting Astronomer, Infrared Telescope Facility.
IRTF is operated by the National Aeronautic and Space Administration}
\altaffiltext{2}{Email: cryan@nmsu.edu}

\begin{abstract}
We present phase-resolved low resolution infrared spectra of the
polar EF Eridani obtained over a period of 2 years with SPEX on the IRTF. The
spectra, covering the wavelength range 0.8 $\leq$ $\lambda$ $\leq$ 2.4 $\mu$m,
are dominated by cyclotron emission at all phases. We use a ``Constant Lambda''
prescription to attempt to model the changing cyclotron features seen in the
spectra.
A single cyclotron emission component with B $\simeq$ 12.6 MG, and a plasma
temperature of kT $\simeq$ 5.0 keV, does a reasonable job in matching the
features seen in the $H$- and $K$-bands, but fails to completely reproduce the 
morphology
shortward of 1.6 $\mu$m. We find that a two component model, where both
components have similar properties, but whose contributions differ with
viewing geometry, provides an excellent fit to the data. We discuss the
implications of our models and compare them with previously published results.
In addition, we show that a cyclotron model with similar properties to those
used for modeling the infrared spectra, but with a field strength of
B = 115 MG, can explain the GALEX observations of EF Eri.
\end{abstract}


\section{Introduction}

Polars, or AM Herculis stars, constitute an important subclass of the
cataclysmic variables (CVs) where the white dwarf primary is highly
magnetic (see Wickramasinghe \& Ferrario 2000 for a review).  Like
non-magnetic CVs, polars are interacting binary systems containing
white dwarf primaries (WD) and low-mass main sequence secondary stars.
The accretion pathway of non-magnetic CVs is well known; matter flows from the
secondary star through the inner-Lagrangian point, free-falling until it
settles into an accretion disk around the primary star. However, in polars the
large magnetic field of the primary WD alters the system characteristics in the
following three ways. First, the magnetosphere deflects material
from it's ballistic trajectory before an accretion disk can form. Secondly,
polars are phase-locked. Dipole-dipole interactions between the primary and
secondary star cause the rotation period of
each star to relax to the orbital period of the system on a relatively short
time-scale. Finally, polars have magnetic fields which usher accreting
matter to a stationary accretion column. These occur at one/both of the
magnetic poles of the WD in the ideal, dipolar case.

As the material is transported to the accretion region, the plasma is ionized
mainly as a result of particle collisions, and from X-ray heating by the
accretion region itself. For large mass accretion rates, a standing
hydrodynamic shock is formed near the white dwarf photosphere, with shock
heights usually several percent of its radius. Downstream from the shock, the
electrons gyrating around the magnetic field emit cyclotron radiation.
Lamb \& Masters (1979) suggested this as the primary cause of the large optical
linear and circular polarization ($\sim$ 10$\%$), from which Tapia (1977a) had
deduced a magnetic field strength of $\sim$ 200 MG for AM Her (now known 
to have a primary field strength of 12 $\pm$ 0.5 MG; Kafka et al. 2006), a 
result which firmly
established the high magnetic field strengths of these objects. Much of the
cyclotron radiation emitted near the shock travels downward and is absorbed by
the stellar photosphere, where it can be reprocessed. Low Accretion Rate Polars
(LARPs) are defined as polars in which $\dot{m}$ $\leq$ 1 g s$^{-1}$ $cm^{-2}$
over the accretion spot, corresponding to
$\sim\dot{M}$ $\leq$ 1$\times$10$^{-14}$ M$_{\odot}$ yr$^{-1}$ at L$_{1}$.
In such systems the timescale required to effectively cool the particle stream
from its free-fall energy is less than a mean free path through the stellar
atmosphere. This disrupts the formation of the hydrodynamic shock, and causes
radiation to be emitted at temperatures lower than those predicted by strong
shock models (c.f., Fabian et al. 1976), directly depositing the particle
stream on to the stellar photosphere in the ``bombardment scenario''
(Kuijpers \& Pringle, 1982; Thompson \& Cawthorne, 1987).

EF Eri is an ultra-short period polar (P$_{\rm orb}$ = 81 minutes), that
has remained an object of considerable interest since entering an extended low
state in 1997 (Wheatley \& Ramsay 1998). Recent SMARTS data indicate that it
has stayed at $V$ = 18.2 $\pm$ 0.1 since that time with only one exception: a
brief flare of 2.5 mag on 2006 March 05 (Howell, S., private communication).
EF Eri quickly returned to its low state two weeks later. Aside from this 
isolated event, EF Eri has continued in a nearly identical low optical state 
for at least the last 10 years.

Ferrario et al. (1996) detected cyclotron harmonics in an orbitally
phase-averaged near-infrared spectrum of EF Eri in its high state, and found
that a model with two different magnetic field strengths ($B_{1}$ = 16.5 MG,
$B_{2}$ = 21.0 MG) best fit their data.
Using Zeeman splitting of the Balmer lines, Wheatley \& Ramsay (1998) derived a
field strength of 13 MG. More recently Reinsch et al. (2003), and Beuermann
et al. (2007) employed Zeeman tomography to map the
field structure over the surface of the white dwarf. In the highest order
multipole expansion considered ($l_{\rm max}$ = 5) EF Eri was found to possess
complex zones and/or bands of varying field strength containing regions of low
field strength (B $\sim$ 10 - 15 MG), as well as a well-defined $\sim$ 100 MG 
pole.

Below we present new phase-resolved spectra of EF Eri in the infrared showing
that large, variable cyclotron features are clearly responsible for the near-IR
photometric variations observed by Harrison et al. 2004 (henceforth, H04).
In the next section we describe our observations, in section 3 we fit
these data with a changing cyclotron model, discuss our results
in section 4, and draw our conclusions in section 5.

\section {Observations}

EF Eri was observed using SPEX on the Infrared Telescope Facility (IRTF) on
the nights of 2004 August 17, and 2007 January 14. SPEX was used in
low-resolution ``prism'' mode with a 0.3'' x 15'' slit. To remove background,
we nodded EF Eri along the slit in an ABBA pattern. In its low-resolution mode
SPEX produces R $\sim$ 250 spectra, with short enough exposure times to
obtain phase-resolved spectra of polars with $K \leq 16.0$. For the two
epochs of observation, we used 240 second and 360 second exposure times,
respectively, which were then median combined with 2-3 other spectra to
allow for cosmic-ray removal and to increase the S/N ratio. The spectra were
reduced using the SPEXTOOL package (Vacca et al. 2003). A telluric correction
was applied using an A0V star of similar airmass to EF Eri. The stacked series
of phase-resolved spectra are shown in Fig. 1 covering 0.8 to 2.4
$ \mu m$. The observed spectra are dominated by a series of broad features 
which can only be construed as cyclotron features. Despite the existence of a 
new spectroscopic ephemeris for EF Eri (Howell et al. 2006a) we have phased 
our spectra using the photometric ephemeris of Bailey et al. (1982) to enable 
direct comparison with previously published lightcurves.

During the 2004 August observations, a telescope issue resulted in only partial
phase coverage. Thus, we returned in 2007 January to obtain full orbital
coverage of EF Eri. There appears to have been little change in the
phase-dependent morphology of the spectra in the intervening three years.
Given the narrow slit width (0.3'') of SPEX, flux calibration of the spectra
is uncertain. Thus, we have used the 2001 $K$-band light curve fluxes
presented in H04 to flux calibrate the spectra in this bandpass at each phase.
H04 also showed that the WD contributes up to 67$\%$ of the $J$-band flux near
$J$-band minimum. We found that it was necessary to account for this component
in the modeling process. For the models described below, we approximate the WD
spectrum as a 9750 K blackbody, normalized at 1.00 $\mu m$ to a flux of 2.41
$\times$ 10$^{\rm -13}$ erg s$^{\rm -1}$ cm$^{\rm -2}$ (Schwope et al. 2007; 
hereafter ``S07'').

\section{Cyclotron Modeling}

As is evident from the infrared spectra presented in Fig. 1, the changing
cyclotron emission is the primary cause of the near-infrared photometric
variations of EF Eri. The spacing of the harmonics seen in these data are
consistent with field strengths near B = 13 MG. Given the coverage of at least
five harmonics, we can attempt to derive the conditions that give rise to this
emission, and attempt to localize the accretion region(s) on the white dwarf.
For this purpose we will employ a simple cyclotron modeling code.

\subsection{An Introduction to Constant Lambda Modeling}

For magnetic field strengths displayed by the primary stars of polars, the 
cyclotron spectrum transitions from optically thick to thin at optical and/or 
near-IR wavelengths, making complex radiative transfer calculations necessary 
to model the emitted spectrum. As we shall see, ``constant lambda'' (CL) 
models provide a straightforward proscription by which the emerging cyclotron
spectrum can be calculated using four parameters, all tied to global properties
at the accretion spot.

In the accretion column, the plasma is dispersive and birefringent.
Ramatay (1969) showed that in the large Faraday rotation
limit of such a plasma, $\phi >> 1$, where $\phi=\frac{e^{3}\lambda^{2}}
{2\pi m^{2}c^{4}} \int_{0}^{s}n_{e} \vec B(s) ds$, it is possible to decouple
the radiative transfer into two magnetoionic modes (ordinary and extraordinary). In this case, the radiative transfer equation reduces to $I_{o,e}=I_{RJ}
[1-exp(-\tau_{o,e})]$ where the ${\it o}$ and ${\it e}$ indicate the ordinary
and extraordinary modes, respectively. The total intensity is taken to be the
sum of the ordinary and extraordinary modes.

The optical depth, $\tau$,  can be parametrized in terms of a
dimensionless optical depth (or ``size'') parameter $\Lambda $: $\tau_{o,e} =
\Lambda \phi_{o,e}$ where $\Lambda= l\omega^{2}_{p}/(c\omega_{c})$. $l$ is the
path length through the plasma, $\omega_{p}$ is the plasma frequency,
$\omega_{p}= (4\pi Ne^{2}/m)^{1/2}$, and $\omega_{c}$ is the frequency of the
cyclotron fundamental, $\omega_{c}=eB/mc $, with $m$ being the
relativistic mass of the gyrating particles. In the above formulation, the
dimensionless absorption coefficient $\phi_{o,e}$ is dependent on B, the
magnetic field strength, $\Theta$, the viewing angle with respect to the
magnetic field, and T, the isothermal temperature of the 
emitting slab. By
integrating the emissivity of the gyrating electrons over an assumed
relativistic Maxwellian distribution, Chanmugam \& Dulk (1981; CD81) produced a
general formulation of the cyclotron absorption coefficients.

These CL models, so called because the emergent radiation is assumed to be due
to a single path through a slab of
uniform optical depth parameter, $\Lambda$,  were used with great success in
the past (see Wickramasinghe and Meggitt, 1985a; Schwope, 1990).
CL cyclotron models depend on four distinct global variables mentioned 
previously: B, T, $\Theta$, and log($\Lambda$). Because altering those four 
parameters causes complex, quasi-degenerate changes in the spectra produced, 
we briefly examine the influence of each parameter on the emerging cyclotron 
spectra below. These effects are detailed in Fig. 2.

The magnetic field strength  is the most independent of the four
parameters used, as increasing B merely shifts the position of each harmonic
blueward. Increasing the plasma temperature has two main
effects. Primarily, it causes the harmonics to grow. But as they bleed upwards
they eventually ``saturate'' at the Rayleigh-Jeans limit (proportional
to $\lambda^{-4}$), effectively becoming a black-body source. Because the
lower harmonics saturate first, higher harmonics contribute ever more of the
total flux with increasing temperature. Also, because the humps are the
product of an ensemble of electrons emitting over a temperature distribution,
higher temperatures increasingly populate the wings of the relativistic
Maxwellian, broadening each harmonic as a result. Unfortunately, each of these
affects can also be reproduced by increasing log($\Lambda$), the
``size parameter'' of the system. Because the temperature does produce a small
systematic shift in the position of the harmonics, it is possible in theory to
decouple the affects of temperature and the size parameter. However, in
practice, identifying this shift is difficult.  Finally, varying the viewing
angle changes the spectrum in two ways.  First, viewing angles near
90$^{\circ}$ produce ``peaky'' harmonics with the higher orders pumped up.
Also, as the viewing angle decreases from 90$^{\circ}$, the harmonics
shift blueward. This viewing angle induced shift then creates a periodic
motion in the position of the harmonics over the orbit as the viewing angle
ranges between its minimum and maximum values.

\subsection{One-Component Cyclotron Models}

The $JHK$ light curves of EF Eri presented in H04 were phased using the
Bailey et al. (1982) photometric ephemeris, which was also
used to phase the current work. The $H$- and $K$-band lightcurves
had minima at $\phi$ = 0.9 and amplitudes of 0.7 and 0.8 mag, respectively.
Remarkably, the $J$-band lightcurve was nearly anti-phased to $H$ and $K$
bands, with a minimum occurring at $\phi$ = 0.4. These broadband variations
can be deduced from the changing morphology of the $JHK$ spectra seen in Fig.
1. We will show that cyclotron emission is the dominant source for EF Eri's
infrared variations.

Qualitatively, the $H$ and $K$ variations are due to a gradual increase in the
size of the cyclotron harmonics in those bands from $\phi$ = 0.00 to
0.50, and their subsequent decline thereafter. While the gradual growth in the
cyclotron features is readily apparent in the $H$-band, in the $K$-band,
the arbitrary flux offset and relative flatness of the mostly optically-thick
($n$ = 4) cyclotron harmonic conspire to make its growth less obvious in
Fig. 1.  Nevertheless, the $K$-band flux (averaged over the range 2.02 to 2.44
$\mu$m) increases from $\lambda F_{\lambda}$= 5.42$\times$10$^{-13}$ to
1.06$\times$10$^{-12}$ erg s$^{-1}$ cm$^{-2}$ over the interval
$\phi$ = 0.01 to 0.45. The source of the variation in the $J$-Band is
also apparent in the SPEX data. From $\phi$ = 0.79 to 0.29 the spectrum in the
$J$-band is flat, with a slight photometric upturn at its shortest wavelengths.
Over the interval $\phi$ = 0.34 to 0.63, however, the $J$-band slope is
qualitatively steeper, causing the brief $J$-band minimum seen in the light
curve.

We produced cyclotron models using a CL code first developed by Schwope et
al. (1990), co-adding the cyclotron spectrum with that of a 9750 K blackbody,
normalized to the S07 WD flux. Because of the telluric features,
and the low signal-to-noise of our spectra, an automated least-squares
minimization procedure was not
employed. Rather, we used published constraints for each parameter from the
literature as well as our own modeling experience to hone in on the
``reasonable'' section of possible parameter space.
First-order formulations for the wavelength dependence of the observed
cyclotron humps with field strength (c.f. Wickramsinghe \& Ferrario, 2000)
imply fields of order 14 MG for EF Eri. This value is similar to the
primary field strengths cited previously. We produced models covering the
range of 10 $\leq$
B $\leq$ 21 MG. Constraining the plasma temperature was more difficult, and
required us to investigate a large range for this variable. In the end, we
generated models from a lower limit of kT = 1 keV, up to the published
high-state temperature of kT= 14.0 keV (Done et al., 1995).  For both low
angles, and low values of log($\Lambda$), cyclotron harmonics become indistinct
with hardly any flux in the higher ($n$ $\geq$ 4) harmonics. Based on this
knowledge, and the fact that multiple distinct cyclotron features are obvious
in our data, we only fit models with $\Theta > 35^{\circ}$ and
1.5 $\leq$ log($\Lambda$) $\leq$ 7.5.

Our model spectra are shown in Fig. 1, with their phase-dependent parameters
listed in Table 1. The magnetic field strength varied slightly, with
a mean value of 12.65 MG and a range of $\pm$ 0.15 MG. The
plasma temperature was found to have a nearly constant value of kT = 4.5 keV,
increasing to 5.5 keV near $\phi$ = 0.63.
Meanwhile, $\Lambda$ changed by a factor of 2, varying
between 5.4 $\leq$ log($\Lambda$) $\leq$ 5.7, with the minimum occurring near
phase 0.60, and the maximum near $\phi$ = 0.30.  Finally, the viewing angle has
a maximum of 66$^{\circ}$  at $\phi$ = 0.95, and a minimum at $\phi$ = 0.34,
where $\Theta$ = 51$^{\circ}$. We list reduced chi-squared values, 
$\chi^{2}_{\nu}$, for our fits in Table 1.

Because of the relative simplicity of this one
component model, extracting the system geometry is straightforward. We take
the system inclination to be $i$ = 58$^{\circ}$, the mean value of the viewing
angle in our models. The modulation to the viewing angle over an orbit is
due to the magnetic colatitude of the accretion spot rotating into and away
from our line-of-sight. We find that a magnetic colatitude of 6$^{\circ}$
produced a series of viewing angles consistent with our data.

\subsection{Two Component Cyclotron Models}

The single component CL models provided excellent fits to the spectra for two
thirds of the orbit. But, as shown by the values of $\chi^{2}_{\nu}$ in Table 
1, failed to provide the same quality fits near $\phi$ = 0.5. This is due to
the deviation of the spectra from our models first seen near $\phi$ =
0.34, predominately affecting the regions shortward of 1.6 $\mu$m. To better
match the
evolving spectra, we constructed two-component models, created by coadding two
cyclotron models together, each with independent values of B, T, $\Theta$, and
$\Lambda$. In summing these two models we continually adjust the relative
contribution of the two cyclotron components, normalized to the observed
spectrum in the $K$-band. Table 2 lists the best-fit values for B, T, $\Theta$,
and $\Lambda$, as well as the flux-weighting factor (``F'') at each orbital
phase. We also include the resulting $\chi^{2}_{\nu}$ values for these fits.

Figure 3 depicts the contribution from each of the cyclotron components, an
optically thinner component (hereafter ``thin'') and an optically thicker
(``thick'') component. As shown in table 2, the thin component is cooler, with 
an average temperature of approximately 1 keV lower than the thick component.
The final spectra (shown in green) are the composites
of both cyclotron models added to the normalized WD discussed previously.
The two component models do a better job of explaining the changing
morphology of the spectra with substantially improved $\chi^{2}_{\nu}$ values
in the phase interval 0.34 $\leq$ $\phi$ $\leq$ 0.63.

In Fig.4 we show how the properties of each of the two cyclotron
components change over an orbit, including synthetic lightcurves derived from
our two component fits. The changes in the $J$-band spectra are explained by
the fact that near phase 0.00 the thin and thick components are approximately
equal. Between $\phi$ = 0.11 and 0.56, the flux from the thin
component slowly declines, while that from the thick component increases by
a factor of six. Our models do not perfectly reproduce the $J$-band light
curve. As shown in Fig. 4, the light curve of EF Eri in the $J$-band is
complex, with large photometric variations on timescales as
short as $\Delta \phi$ $\leq$ 0.10. To obtain a reasonable S/N, we have had to
median a number of spectra together, and this process limits our ability to
detect variations on time scales of $\Delta \phi$ $\lesssim$ 0.15. The main
discrepancy between the light curve derived from the models, and that
derived from photometry occurs at $\phi$ = 0.46. The origin for this
difference cannot be identified given the low temporal resolution of our
medianed spectra.

Our two component models provide a much better explanation to the origin of
the changes see in the light curves of the $H$-, and $K$-bands.
In the $H$-band, the thin component dominates near $\phi$ = 0.00. Over the
orbit, this component stays roughly constant, while the flux from the
thick component peaks sharply near $\phi$ = 0.40. Simultaneously, the
contribution from the thin component reaches a minimum as the thick component
peaks. These changes, working in concert, explain the flat $H$-band maxima. The
morphology in the $K$-band is similar to the $H$-band, but with a smaller
contribution from the thick component. Both the thick and thin components
produce fluxes that increase through the first half of the orbit, and then
decline afterward. The more rapid decline of the thick component at
$\phi$ = 0.63 reproduces the light curve feature seen at that phase.

Table 2 summarizes the model parameters for the two component model fits over
an entire orbital cycle. The thick component has a nearly constant magnetic
field strength (12.6 MG) and temperature ($\simeq$ 6.0 keV). The
viewing angle of this component changes from 60$^{\circ}$ to 56$^{\circ}$ over 
the orbit.
Log($\Lambda$), meanwhile, slowly increases for the first half of the orbit
after which it declines. The thin component, by contrast, has a
slightly cooler temperature $\simeq$ 5.0 keV, with a magnetic field strength
which  varies between 12.5 and 12.9 MG. The viewing angle of this component
varies between 64$^{\circ}$ and 53$^{\circ}$. Finally, log($\Lambda$) of this
component declines slowly from 5.5 at $\phi$ = 0.01 to 4.9 at $\phi$ = 0.45.
Afterward, the trend is reversed, reaching log($\Lambda$) = 5.6 at $\phi$ =
0.99. 

\subsection{GALEX Observations of EF Eri}

Szkody et al. (2006) presented GALEX observations of EF Eri. The results
were somewhat surprising, showing significant ultraviolet emission that was
highly variable. In fact, the amplitude of the variability in the FUV bandpass
was nearly identical to that seen in the $H$- and $K$-band light curves.
Szkody et al. attempted to model the combined FUV, NUV and $V$-band light
curves as the sum of a WD + hotspot. While such models could explain both
the SED and UV light curves, they were unable to simultaneously explain the
$V$-band light curve without invoking unusual limb darkening laws.
More recently, S07 has shown that inclusion of a proper WD atmosphere models
produced much better fits to the multi-wavelength light curves.

Given the similar amplitudes of the UV and IR variability, and nearly
identical values of L$_{bol}$(IR) and L$_{bol}$(GALEX), could cyclotron
emission be partly/mostly responsible for the variations seen in the UV light
curves? Obviously, producing significant cyclotron flux in the UV requires
much higher field strengths than used for modeling our near-IR data. But 
model tomographic maps of Beuermann et al. (2007), indicate that
regions of very high field-strength appear to be present in EF Eri.

To explore the possibility of UV cyclotron emission in EF Eri, we have used 
the same parameters for the cyclotron emission as found in 
the one-component models for our SPEX data, except that we adjusted the 
magnetic field strength to allow for UV cyclotron emission. While the value of
model parameters is most likely different in the UV than in the near-IR, we 
have few constraints on the possible emission from this high field component, 
and intend this merely a starting place for modeling work.
Furthermore, because of the discrepancy between the spectrum of a blackbody
and an actual white dwarf atmosphere in the UV/optical, we have replaced the
our 9750 K blackbody with a synthetic white dwarf spectrum (T = 9500 K, 
$log$g = 8, solar metalicity, I. Hubeny, private communication). 
Our results are shown in Fig. 5 for the phases of maximum and
minimum UV fluxes ($\phi$ = 0.45 and 0.95, respectively). These models have
identical values of T and $\Theta$ as the models shown in
Fig. 1, but with B = 115 MG. This field strength is similar to the $\simeq$ 
100 MG spot inferred in Beuermann et al. 2007. Such models do an excellent job 
of matching the minimum light spectral energy distribution (SED) from the FUV 
to the $I$-band. However, to account for the FUV photometric flux observed 
during the bright phases, we had to increase log($\Lambda$) from 5.7 to 6.2 
(the value of $\Lambda$ for the faint state was identical to our near-IR 
models). Further insight into the source of the UV variability of EF Eri will 
be gained once phase-resolved spectroscopy at these wavelengths becomes 
available.

\section{Discussion}

The fact that our one component models fail in the $J$-band, despite their
success in the {\it H}- and {\it K}-bands, motivated us to generate two
component fits for EF Eri. These two component models resulted in
substantially better fits at every orbital phase, though they also do not fully
explain the morphology seen in the $J$-band near $\phi$ = 0.5. The magnetic
field strengths we derive are consistent with those previously reported. It is
interesting to note that Ferrario et al. (1996) also required a two component
model to explain their phase-averaged, high-state infrared spectrum of EF Eri.
In that case, however, moderately higher magnetic field strengths were
necessary.

The high plasma temperatures that we require for our models is surprising
given the long-lived low state of EF Eri, and the expectation that the
associated mass accretion rate is very low, possibly in the ``bombardment
scenario'' regime (Kuijpers \& Pringle 1982). But the presence of
the $n$ = 7 harmonic indicates the plasma temperature cannot be extremely low.
Comparison of plasma temperatures derived from X-ray observations for other
polars in low states are in agreement with what we find for EF Eri. For
example, Ramsay et al. (2004) presented data on 16 polars in low-states
($\dot{m}$ below 10$^{-2}$ g s$^{-1}$ cm$^{-2}$), of which 7 were
detected at sufficient levels to allow their plasma temperatures to be
modeled. They derived temperatures that ranged between 1.4 and 5.0 keV,
demonstrating that despite the low values of $\dot{m}$ in these systems,
some polars can maintain moderately high plasma temperatures.

Until this year, only the high-state temperature for EF Eri had been
modeled using X-ray data. Done et al. (1995) used GINGA to fit a Raymond-Smith
spectrum with kT = 14 keV, with a maximum temperature of 25 keV, a value close
to temperatures predicted by strong-shock models. S07 have
recently used XMM-Newton to detect EF Eri in its low state. The spectrum was
fit as a MEKAL plasma with a temperature of 2.8$\pm$1.7 keV. This 
temperature compares reasonably well to those obtained by in Ramsay et al. 
(2004) for other low $\dot{m}$ polars in the X-ray, but is lower than the 
temperatures derived here; possibly because the X-ray emission 
and the cyclotron emission modeled here are generated at different regions of 
the shock. More pertinently, Fischer \& Beuermann (2001) have produced 
normalized temperature and density profiles in the accretion column of polars, 
running through many lines of sight of the cooling plasma to add a layer of 
realism beyond standard CL modeling. For EF Eri, assuming B = 13 MG and
$\dot{m}$ = 1$\times$ 10$^{-2}$ g s$^{-1}$ cm$^{-2}$ they found a maximum
temperature of 7 keV, consistent with our results. CL models may inherently
require slightly higher plasma temperatures than are actually present.
Rousseau et al. (1996) compared cyclotron models that included an ensemble of 
mass accretion rates (each with an associated plasma temperature), to the 
results from a single CL model for UZ For. They found that the CL model 
required temperatures at high end of the range when compared to those seen in 
the multiple accretion rate model, indicating that perhaps CL plasma 
temperatures are systematically hot.

\subsection{ The Accretion Region Geometry}

As discussed earlier, it is simple to explain the single-component
cyclotron model evolved in section 3.2: the cyclotron emission comes
from a region with a magnetic colatitude of 6$^{\circ}$ (with $i$ =
58$^{\circ}$).  But a single cyclotron component did not fully explain the
evolving morphology of the spectra. Thus we added a second cyclotron emission
component. The two component model resulted in substantially
better fits throughout the orbit. We can derive insight into the geometry
of the cyclotron emission from this two component model by analyzing the
modulation of the viewing angle of each component over an
orbital cycle. The viewing angle, $\Theta$, is defined as:
\begin{equation}
\Theta=\mbox{cos}^{-1}(\mbox{cos}(i)\mbox{cos}(b)-\mbox{sin}(i)\mbox{sin}(b)\mbox{cos}(2\pi\phi)),
\end{equation}

\begin{flushleft}
here $i$, is the orbital inclination angle, $b$ is the angle between the
rotation axis and the direction of the $local$ field line at the accretion
region, and $\phi$ is the orbital phase. It is important to note that $b$ can
be different from the magnetic colatitude, $\beta$, which is defined as the
angle between the rotation axis and magnetic axis. For example, in ideal, 
dipolar accretion (which is almost certainly not the case in EF Eri), the 
diverging field lines produce an approximate relationship of, $b$ $\simeq$ 
$\beta + 3/2\alpha$ (see Beuermann et al. 1987), where $\alpha$ is the angular 
distance from the magnetic axis. Due to the large
angular extent of the accretion regions observed for most polars, $b$ can
exhibit significant variations from the core to the edge of the accretion spot.
\end{flushleft}

In our two component model, the viewing angles of the thin and thick components
range from 64$^{\circ}$ to 53$^{\circ}$ and 60$^{\circ}$ to 56$^{\circ}$,
respectively. For the inclination we used $i$ =  58$^{\circ}$, the mean value
of the viewing angle in both the thin and thick components. This value agrees
with the inclination angle derived by Piirola et al. 1987
($i$ = 55$^{\circ}$ $\pm$ 3). We can then fully specify the viewing angle at a 
given
orbital phase once $b$, the field line angle, is determined. In a simple
accretion picture, with a constant magnetic field, the magnetic co-latitude
$\beta$ will be equivalent to the field line angle $b$. While departures of
up to 0.4 MG are evidenced in our models, this amplitude represents only three
percent of the magnetic field strength locally active. As such, we consider the
magnetic field strength to be nominally fixed, and constant. The magnetic
co-latitude, $\beta$, is then just the maximum deviation of viewing angle, 
from the orbital inclination of the system. Thus, the magnetic
co-latitudes of the thin and thick components are 6$^{\circ}$ and 2$^{\circ}$
respectively.

Our accretion geometry is generally consistent with the picture derived from 
high-state X-ray observations. Beuermann et al. (1987) found that the accretion
region of EF Eri resembled an ``X-Ray auroral oval'', showing  an extended, 
diffuse tail
of $\simeq$ 20$^{\circ}$, with a compact nucleated core near the rotation
axis. The difference of course, is that the X-ray emission comes directly
from the shock, while the cyclotron emission is preferentially emitted at
right angles to this feature. In addition, the actual structure of this
feature could have changed dramatically as EF Eri dropped into its low state.
In practice, our two cyclotron component model has taken this extended
and structured shock and reduced it to two discrete spots: the thick component,
which corresponds to the core of the accretion arc and the thin component
which corresponds to the  tail. However, the two cyclotron emission regions
are only separated by four degrees, and it a distinct possibility that the two
components could correspond to local variations of a single accretion spot.

As previously mention, our motivation for running two component models
largely resulted from our inability to explain the excess emission of the
$J$-band near $\phi$ = 0.40. To some extent, these more complex models
were a success. However, we still fail to adequately reproduce the far
blue-end of the spectra, a fact which may necessitate the need for additional
emission components. If cyclotron emission from accretion on to very
high field strengths is actually present in EF Eri, then those components
could introduce an additional contribution to the $J$-band (note that the
$n$ = 1 harmonic for B = 100 MG is at 1.07 $\mu$m).

\subsection{Where is the Secondary Star in EF Eri?}

Above, we discussed a model for the cyclotron emission from EF Eri that appears
to be reasonable, and explains the large amplitude variations seen in its
near-infrared light curves. As a consequence, the cyclotron models leave
little room for emission from the expected low mass secondary star (but note
the small ``excess'' at the red end of the $K$-band seen in many of our
spectra). With these cyclotron models, we can attempt to put additional limits
on the nature of this object.

Beuermann et al.(2000) and H04 have shown that secondary star in EF Eri
must have a spectral type later than M9 to be consistent with its observed
spectral energy distribution. Howell et al. (2006b) have used radial velocity
observations to show that the secondary appears to have a sub-stellar
mass. While the model fits to the observed spectra are not perfect, it is
difficult to have a $significant$ source of additional infrared luminosity
in this system. For example, if the secondary star is an L-dwarf, it must
supply less than 25$\%$ of the $K$-band flux (at $K$-band minimum) to be
consistent with the observed spectrum.

The observed magnitude at light curve minimum is $K$ = 15.65 (H04). Thorstensen
(2003) has published a parallax for EF Eri, with the formal result of d = 163
(+63/-50) pc. Including priors that include the Beuermann et al. (2000) result
for the white dwarf, and the sizable proper motion, they arrive at a lower
distance of 113 (+19, -16) pc. Using the entire possible range in distance
(97 to 226 pc), and assuming the secondary star supplies $\leq$ 25$\%$ of the
$K$-band luminosity, the maximum observed absolute magnitude for the secondary
of EF Eri is in the range 11.9 $\leq$ M$_{\rm K}$ $\leq$ 10.1, completely
consistent with the presence of a brown dwarf secondary.

\section{Conclusion}
We have presented new, phase-resolved low resolution spectra of
EF Eri that demonstrate its near-IR SED is dominated by cyclotron emission.
We have constructed models using a CL prescription that
is reasonable, and can explain the large amplitude variations observed in
its $JHK$ light curves.  In addition, we also show that cyclotron emission may
be responsible for the GALEX observations of EF Eri. Given the complex
magnetic field structure of EF Eri, near-IR observations of higher temporal
cadence would be extremely useful in unraveling the accretion geometry in this
system, but will require an 8-m class telescope.

% The equation environment wil produce a numbered display equation.


%% Included in this acknowledgments section are examples of the
%% AASTeX hypertext markup commands. Use \url without the optional [HREF]
%% argument when you want to print the url directly in the text. Otherwise,
%% use either \url or \anchor, with the HREF as the first argument and the
%% text to be printed in the second.

\acknowledgments

\clearpage
\begin{deluxetable}{lllllll}
\tabletypesize{\scriptsize}
%\rotate
\tablecaption{Parameters for the One Cyclotron Component Models}
\tablewidth{0pt}
\tablehead{
\colhead{Phase} & \colhead{B (MG)} & \colhead{T (keV)} &
\colhead{$\Theta$} &\colhead {$Log(\Lambda)$} &\colhead {$\chi^{2}_{\nu}$} }
\startdata
0.01 & 12.6 & 4.5 & 64.0 & 5.7 & 1.36 \\
0.11 & 12.6 & 4.5 & 62.0 & 5.9 & 1.39 \\
0.23 & 12.6 & 4.5 & 58.0 & 5.6 & 2.06 \\
0.29 & 12.5 & 4.5 & 56.0 & 6.0 & 1.35 \\
0.34 & 12.6 & 4.5 & 51.0 & 5.7 & 2.05 \\
0.45 & 12.6 & 4.5 & 55.0 & 5.7 & 2.58 \\
0.56 & 12.6 & 4.5 & 58.0 & 5.7 & 2.29 \\
0.63 & 12.8 & 5.5 & 58.0 & 5.4 & 2.35 \\
0.79 & 12.8 & 5.0 & 60.0 & 5.5 & 1.36 \\
0.95 & 12.8 & 5.5 & 66.0 & 5.6 & 1.26 \\
0.99 & 12.6 & 5.0 & 66.0 & 5.6 & 1.33 \\
\enddata
\end{deluxetable}

\clearpage

\begin{deluxetable}{llllllllllll}
\tabletypesize{\scriptsize}
%\rotate
\tablecaption{Parameters for the Two Cyclotron Component Models}
\tablewidth{0pt}
\tablehead{\colhead{Phase} & \colhead{$F_{1}$} & \colhead{$F_{2}$} &
\colhead{$B_{1}$ (MG)}  & \colhead{$B_{2}$ (MG)} & \colhead{$T_{1}$ (keV)} &
\colhead{$T_{2}$ (keV)} & \colhead{$\Theta_{1}$} & \colhead{$\Theta_{2}$} &
\colhead {$Log\Lambda_{1}$} & \colhead {$Log\Lambda_{2}$} &
\colhead {$\chi^{2}_{\nu}$}}
\startdata

0.01 & 0.88 & 0.12 & 12.6 & 12.5 & 4.5 & 6.0 & 64.0 & 60.0 & 5.5 & 6.4 & 1.20\\
0.11 & 0.88 & 0.12 & 12.6 & 12.6 & 5.0 & 6.0 & 61.0 & 60.0 & 5.5 & 6.5 & 1.31\\
0.23 & 0.82 & 0.18 & 12.8 & 12.5 & 5.0 & 6.5 & 59.0 & 59.0 & 5.0 & 6.5 & 1.32\\
0.29 & 0.85 & 0.15 & 12.6 & 12.5 & 5.0 & 6.0 & 57.0 & 58.0 & 5.3 & 6.5 & 1.32\\
0.34 & 0.82 & 0.18 & 12.7 & 12.5 & 5.0 & 6.5 & 55.0 & 57.0 & 4.9 & 6.6 & 1.21\\
0.45 & 0.78 & 0.22 & 12.7 & 12.5 & 5.0 & 6.5 & 53.0 & 56.0 & 4.9 & 6.6 & 1.83\\
0.56 & 0.70 & 0.30 & 12.5 & 12.5 & 5.0 & 6.0 & 55.0 & 56.0 & 4.9 & 6.6 & 1.82\\
0.63 & 0.81 & 0.19 & 12.9 & 12.6 & 5.5 & 6.5 & 57.0 & 58.0 & 4.9 & 6.6 & 1.72\\
0.79 & 0.85 & 0.15 & 12.7 & 12.6 & 5.0 & 6.0 & 60.0 & 59.0 & 5.2 & 6.4 & 1.10\\
0.95 & 0.90 & 0.10 & 12.7 & 12.6 & 5.0 & 6.0 & 63.0 & 60.0 & 5.6 & 6.4 & 1.41\\
0.99 & 0.88 & 0.12 & 12.6 & 12.6 & 5.0 & 6.0 & 64.0 & 60.0 & 5.6 & 6.4 & 1.51\\
\enddata
\tablecomments{Subscripts '1' and '2' refer to the thin and thick components
respectively. F$_{1}$ and F$_{2}$ show the relative contribution of each
cyclotron component to the flux at 2.19 microns.}
\end{deluxetable}
\clearpage

{\bf Figure 1.}
One-component model fits to the IRTF/SPEX dataset for EF Eri. At
each phase, the SPEX data are shown in black. 
The best fit cyclotron component is added to a 9750 K blackbody which is 
normalized to match the  1 $\mu$m WD flux in  S07, to yield the composite model
 (green). At each phase, both the SPEX data and models are
offset by a constant increment of 1.0$\times 10^{-12}$ erg s$^{-1}$ cm$^{-2}$ 
from the previous spectrum: the bottom spectrum corresponds to the labeled 
flux. The orbital phases are printed on the far right margin for 17 August 
2004 (black) and 14 January 2007 (blue). Above the bottommost spectrum the 
cyclotron harmonic numbers are indicated.

{\bf Figure 2.}
Variations on a Theme: Cyclotron Modeling. In the four following boxes
 a nominal cyclotron model with B = 30 MG, kT = 5 keV, $\Theta$ = 80, and
$log(\Lambda)$ = 3 is varied. Increasing values for each parameter proceed
upward in each box. In panel (a) the magnetic field, B, changes from
10 to 35 MG with 5 MG steps. (b) we alter the temperature from 1 to
25 keV in steps of 5 keV. (c) $log(\Lambda)$ is varied from 1 to 6 in
steps of 1. (d) $\Theta$ changes from 30$^{\circ}$ to 80$^{\circ}$, with steps
of 10$^{\circ}$. In each of the model sets we normalize the harmonics relative
to each other. The absolute fluxes are arbitrary.

{\bf Figure 3.}
Two-component fits to the IRTF/SPEX spectra(black), showing the thin (red)
and thick (blue) cyclotron components. (a) Phases 0.00 to 0.34.
(b) Phases 0.45 to 0.99. The green line is the composite of both cyclotron
models and the normalized 9750 K blackbody 
(cyan shown plotted for $\phi$ = 0.01). All spectra are sequentially offset 
by $\lambda F_{\lambda}$ = 1.5$\times 10^{-12}$ erg s$^{-1}$ cm$^{-2}$. 

{\bf Figure 4.}
(a) Orbital variations of relative parameters in the 2-component models.
The top panel shows the H04 $J$-Band lightcurve. The second panel shows 
a synthetic lightcurve constructed by integrating the model spectra through 
the $J$ bandpass for both the thin (red) and thick (blue) 
components as well as their composite (green). Panels 3-5 document how kT, 
$\theta$, and log($\lambda$) change from both the thin and thick components 
over the orbit. The solid triangles are plotted for phases which correspond
to those listed in Table 2.

(b) The $H$ and $K$ band lightcurves as shown in  Fig. 4a.

{\bf Figure 5.}
GALEX and optical photometry from Szkody et al. 2006. The points in
red are at the UV/Optical minimum ($\phi=0.38$).  Likewise, in blue
are the data from the maximum ($\phi=0.85$). The cyclotron + WD models (also
in blue and red) shown are nearly identical to the corresponding IR model at
similar phase to the GALEX observations, but now with B = 115 MG.

\clearpage

\begin{figure}
\epsscale{0.80}
\plotone{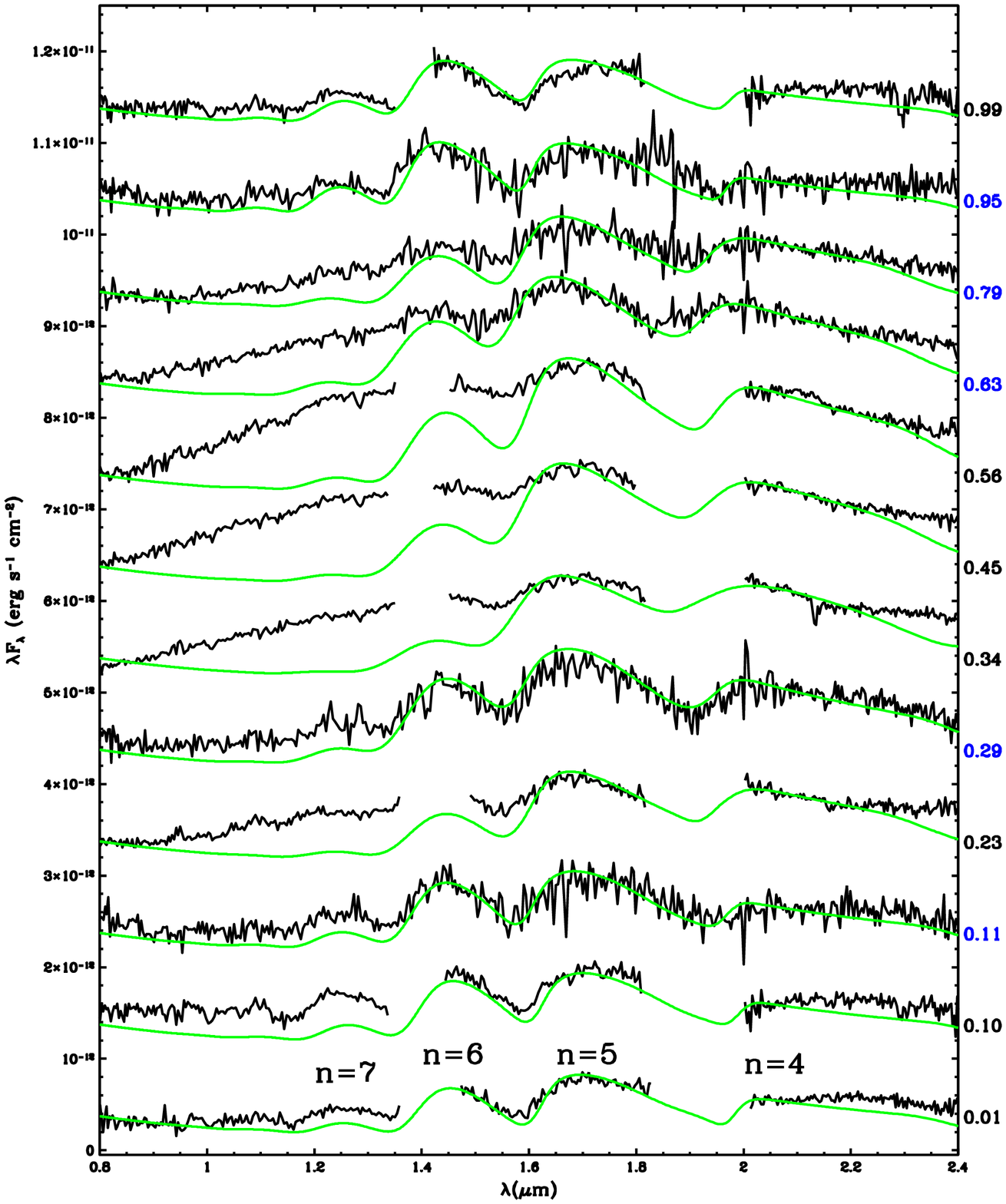}
\end{figure}
\begin{figure}
\epsscale{0.80}
\plotone{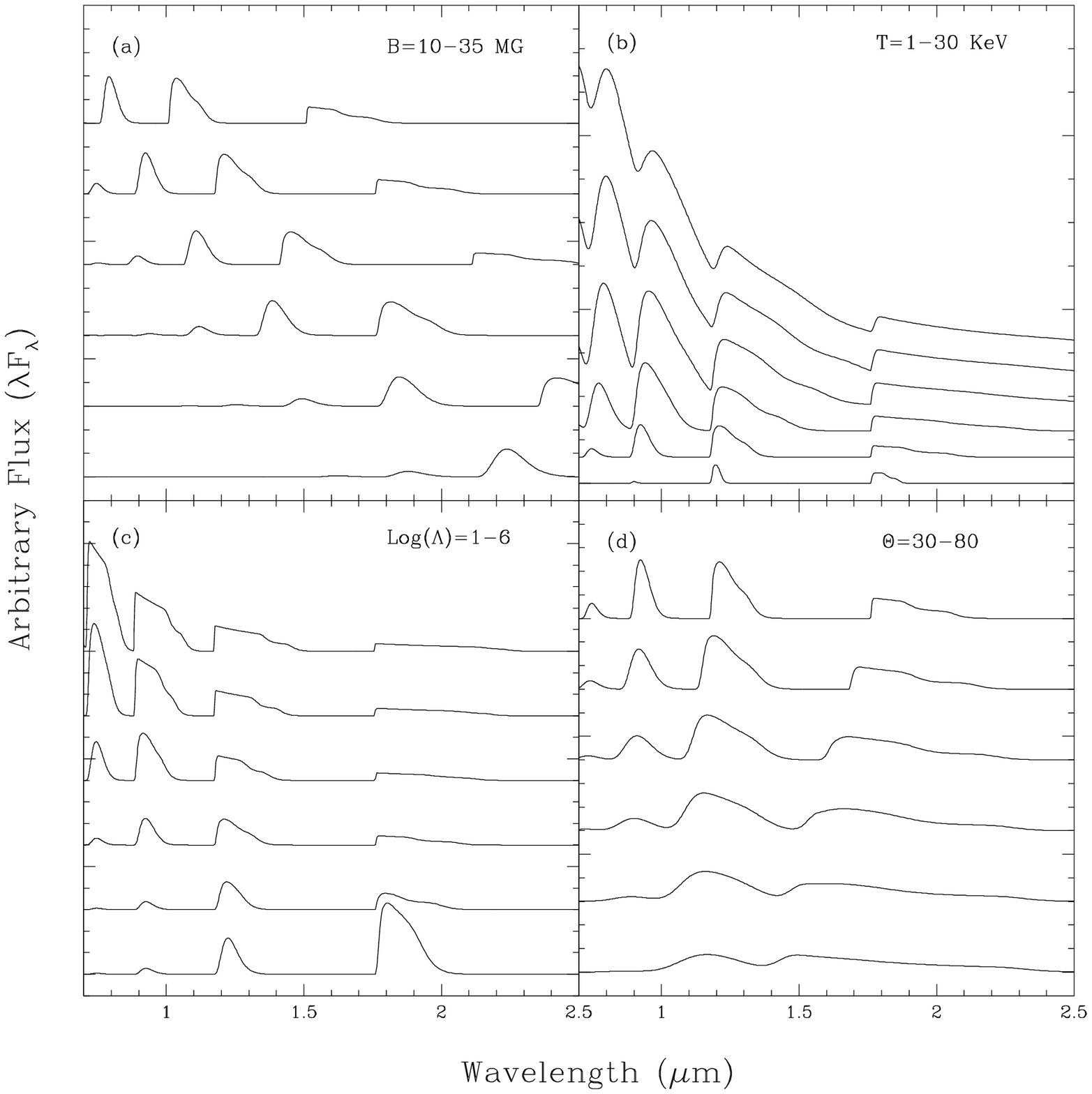}
\end{figure}
\begin{figure}
\epsscale{0.80}
\plotone{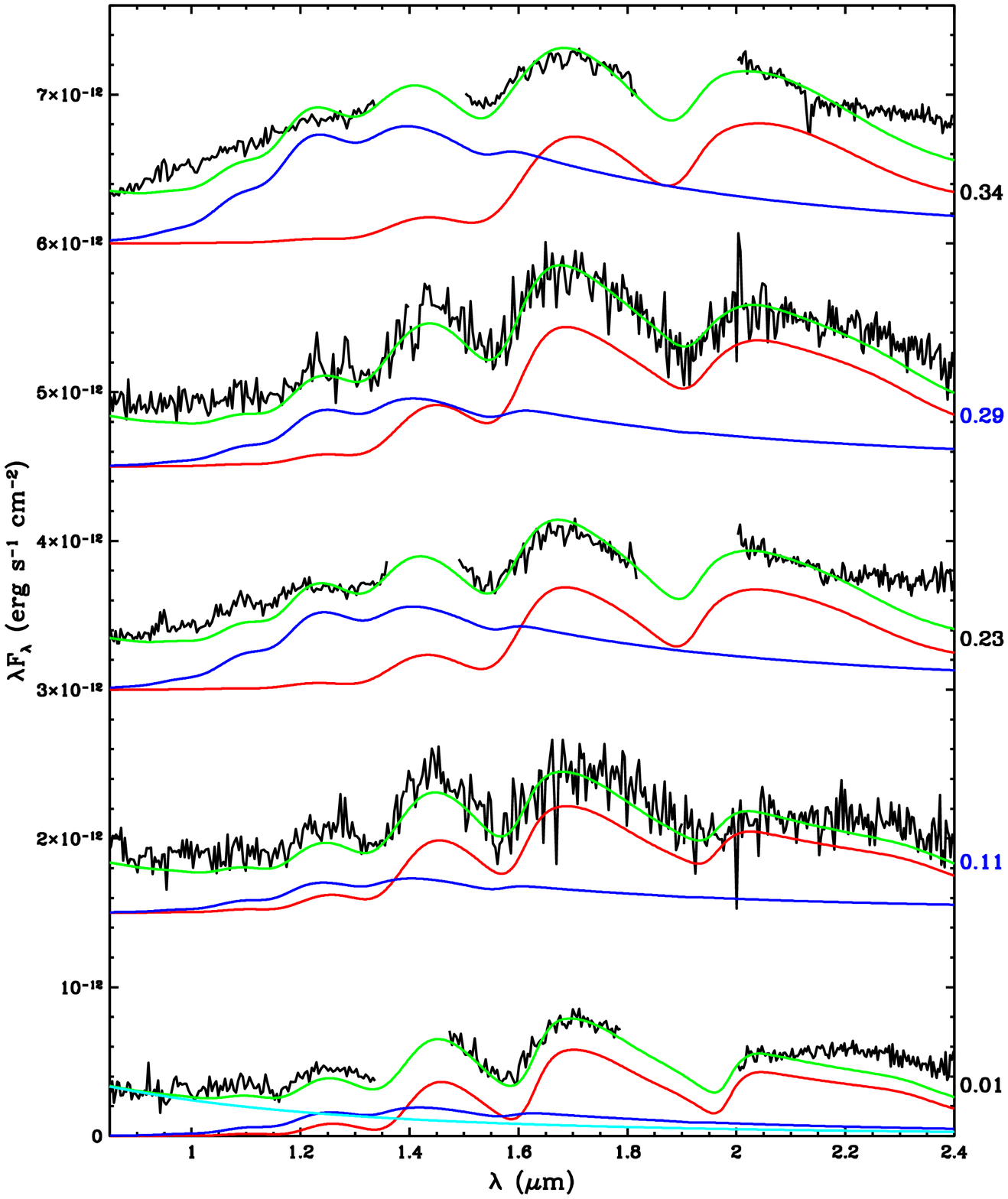}
\end{figure}
\begin{figure}
\epsscale{0.80}
\plotone{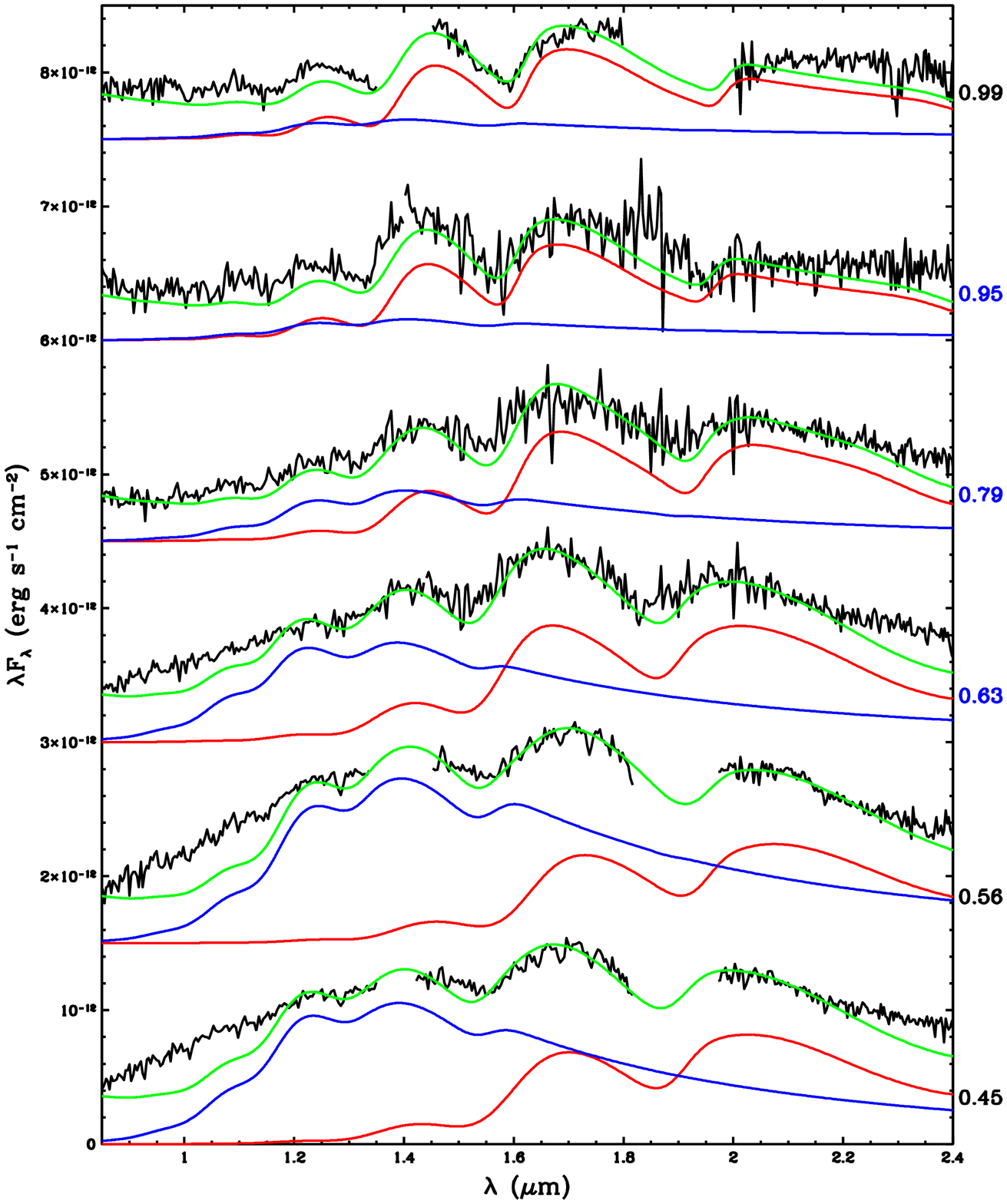}
\end{figure}
\begin{figure}
\epsscale{0.80}
\plotone{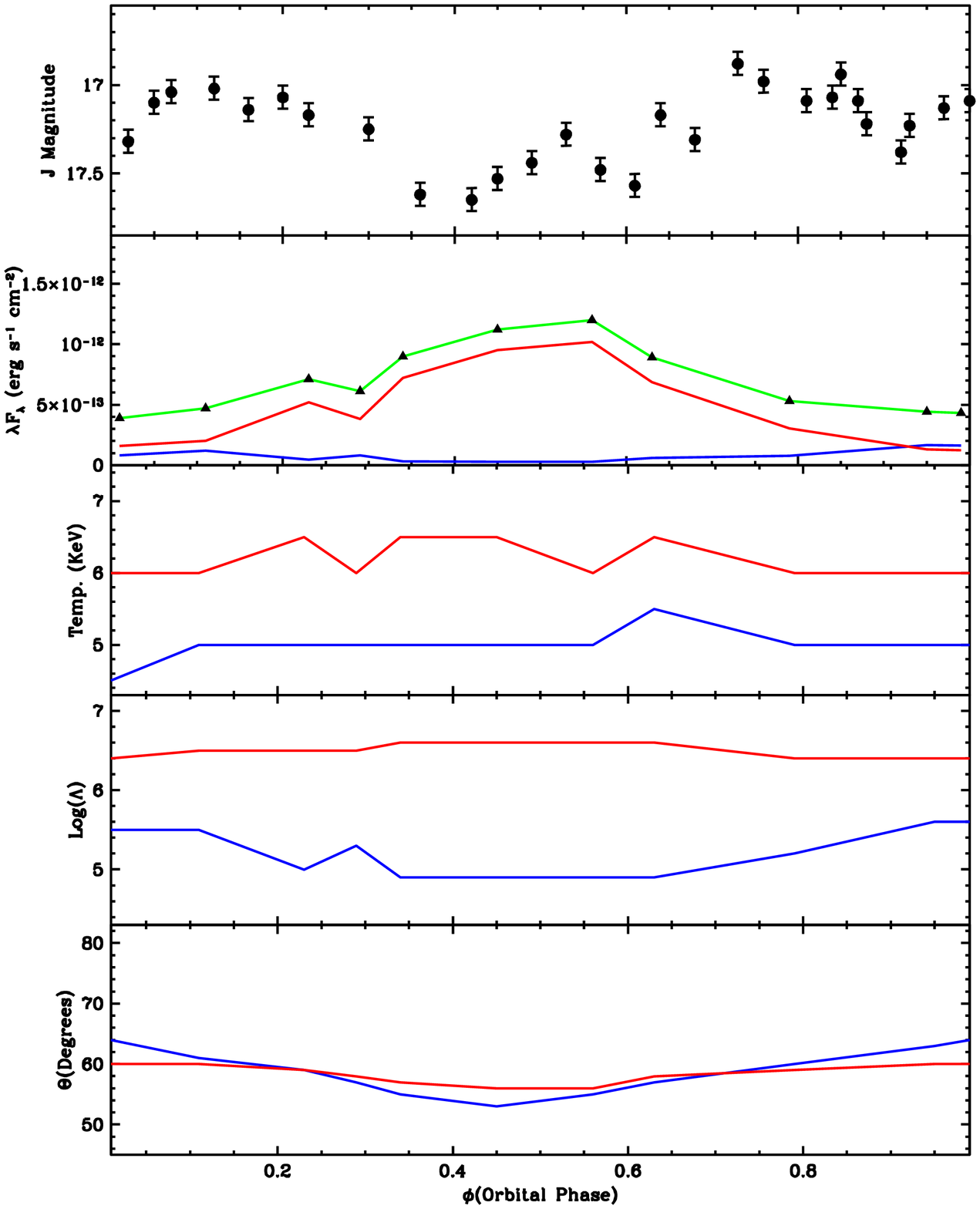}
\end{figure}
\begin{figure}
\epsscale{0.80}
\plotone{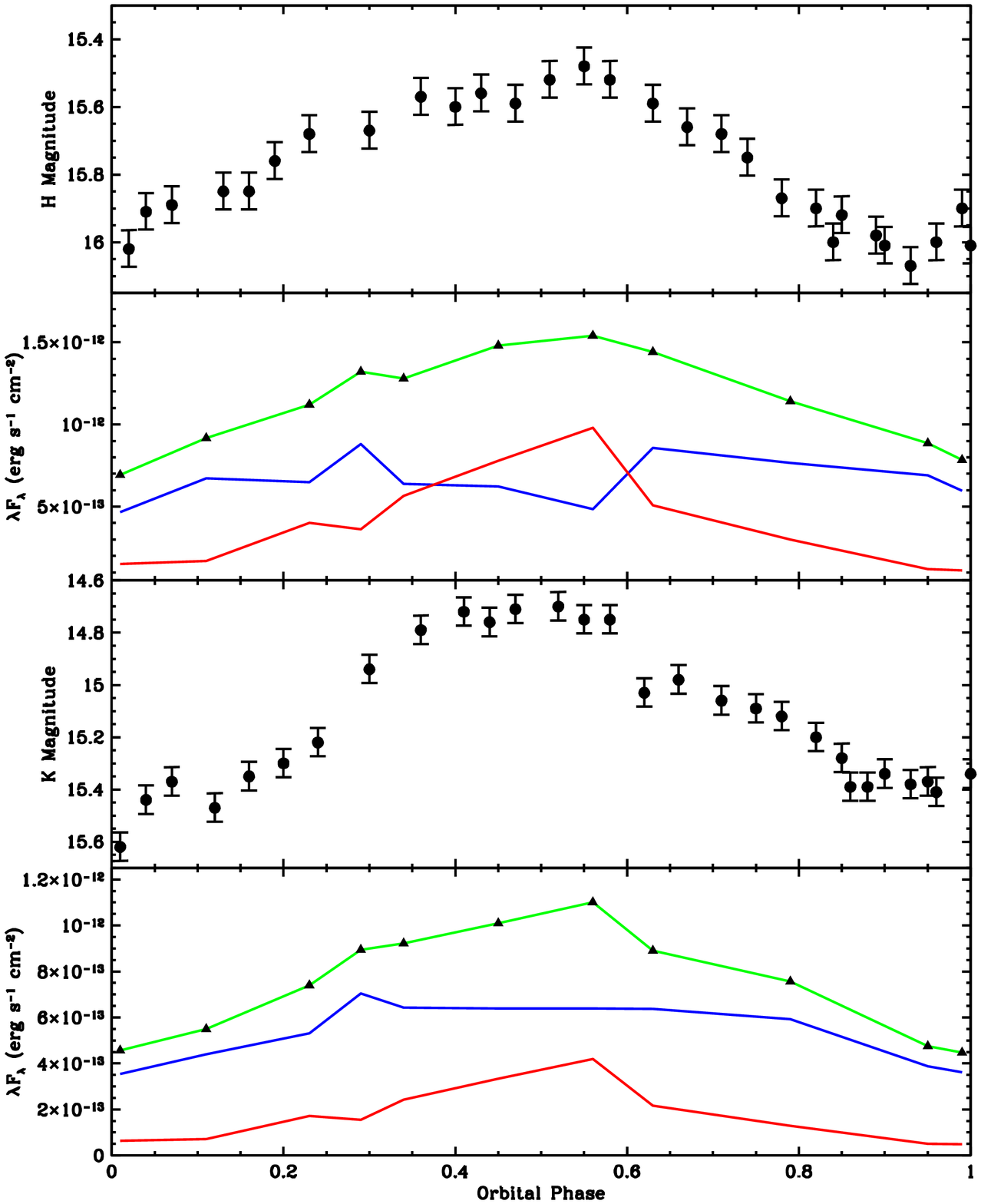}
\end{figure}
\begin{figure}
\epsscale{0.80}
\plotone{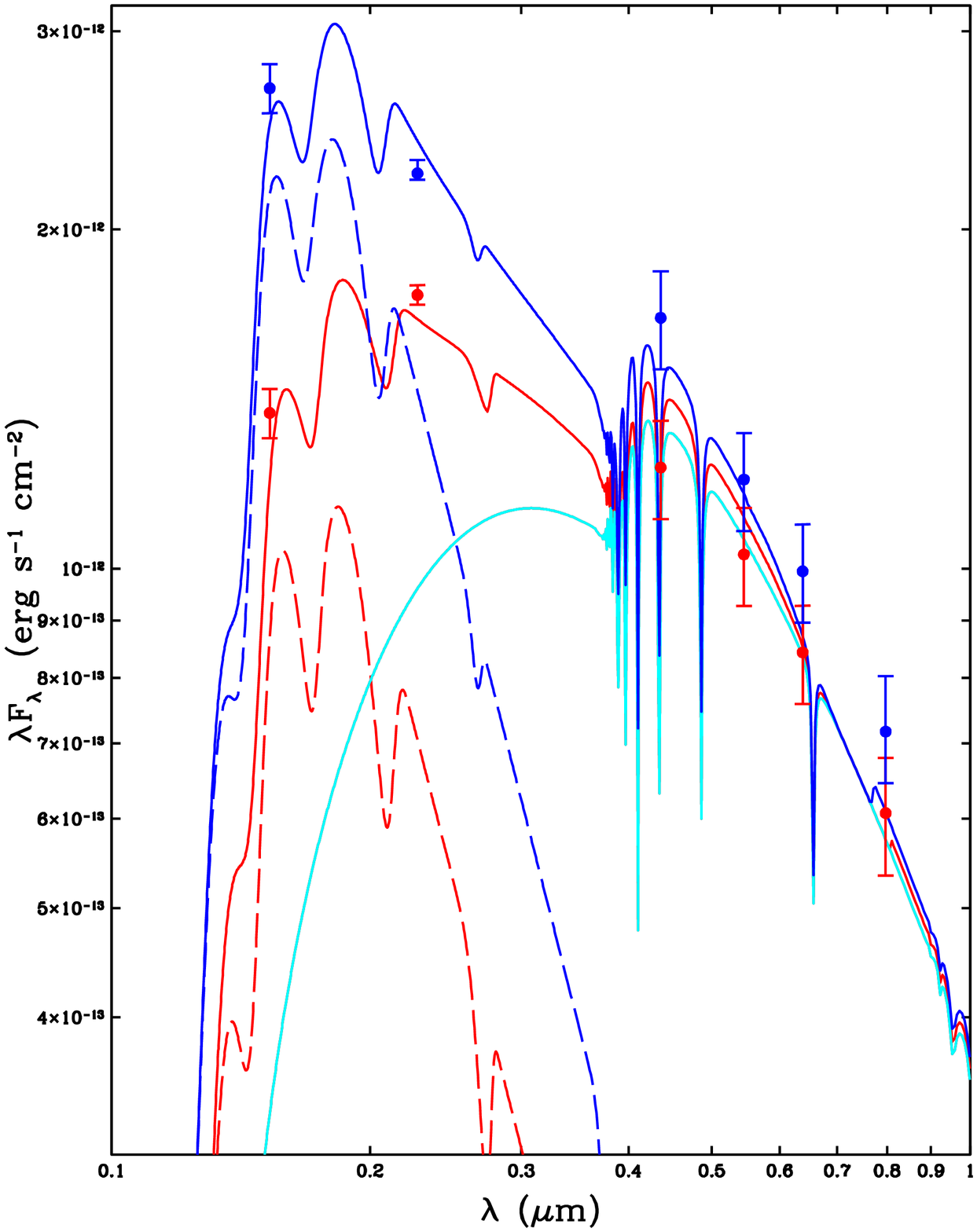}
\end{figure}


\begin{thebibliography}{}
\bibitem[Bailey et al.(1982)]{Bailey82} Bailey, J., Hough, J.H., 
Axon, D.J., Gatley, I., Lee, T.J., Berriman, G., Szkody, P., Stokes, G., 
1982, \mnras, 199, 801
\bibitem[Beuermann, K. et al. (1987)]{Beuermann87} Beuermann, K., Stella, L. 
Paterson, J., 1987, \apj, 316, 360
\bibitem[Beuermann, K. et al. (2000)]{Beuermann00} Beuermann, K., Wheatley, P.,
Ramsay, G., Euchner, F., Gansicke, B. T., 2000, \aap, 354L, 49
\bibitem[Beuermann, K. et al. (2007)]{Beuermann07} Beuermann, K., Euchner, F., 
Reinsch, K., Jordan, S., Gansicke, B. T., 2007, \aap, 463, 647
\bibitem[Chanmugam, G. and G.A. Dulk (1981)]{Chanmugam81} Chanmugam, G., Dulk,
D.A., 1981, \apj, 244, 569
\bibitem[Done, C., et al. (1995)]{Done95} Done, C., Osborne, J.P., 
Beardmore, A.P. 1995, \mnras, 276, 483
\bibitem[Fabian, A.C. et al. (1976)]{Fabian76} Fabian, A.C., Pringle, J.E., 
Rees, M.J., 1976, \mnras, 175, 43
\bibitem[Ferrario, L. et al. (1996)]{Ferrario96} Ferrario, L., Bailey, J., 
Wickramasinghe, D.T., 1996, \mnras, 282, 218
\bibitem[Fischer, A \& K. Beuermann (2001)]{Fischer01} Fischer. A., Beuermann,
K., 2001, \aap, 373, 211
\bibitem[Harrison et al.(2003)]{Harrison03} Harrison, T. E., Howell, S.B., 
Huber, M.E., Osborne, H. L., Holtzman, J. A., Cash, J.L., Gelino, D.M., 
2003, \aj, 125, 2609
\bibitem[Harrison et al.(2004)]{Harrison04} Harrison, T. E., Howell, S.B., 
Szkody, P., Homeier, D., Johnson, J.J., Osborne, H.L., 2004, \aj, 614, 947
\bibitem[Harrison et al.(2007)]{Harrison07} Harrison, T. E., Campbell, R.K., 
Howell, S.B., Cordova, F.A., Schwope, A.D., \apj, 656, 444
\bibitem[Howell, S.B. et al.(2006)]{Howell06a} Howell, S.B., Walter, F.M., 
Harrison, T.E., Huber, M.E., Becker, R.H., White, R.L., 2006a, \apj, 652, 709
\bibitem[Howell, S.B. et al.(2006)]{Howell06b} Howell, S., Harrison, T.E., 
Campbell, R.K., Cordova, F.A., Szkody, P., 2006b, \aj, 131, 2216
\bibitem[Kafka, S. et al.(2006)]{Kafka06} Kafka, S., Honeykutt, R.K, Howell, 
S.B., 2006, \aj, 131, 2673
\bibitem[Kuijpers, J. \& Pringle, J.E. (1982)]{Kuijpers82} Kuijpers, J., 
Pringle, J.E., 1982, \aap, 114, 4
\bibitem[Lamb, D.Q., \&  Masters, A.R. (1979)]{Lamb79} Lamb, D.Q., Masters, 
A.R, 1979, \aj, 234, 117
\bibitem[Meggitt, S.M.A., \& D.T. Wickramasinghe (1989)]{Meggitt89} Meggitt,
S.M.A., Wickramasinghe, D.T., 1988, \mnras, 236, 31
\bibitem[Piirola, V. et al. (1987)]{Piirola87} Piirola, V., Coye, G.V., 
Reiz, A., 1987. \aap, 186, 120
\bibitem[Ramatay, R. (1969)]{Ramatay69} Ramatay, R., 1969, \apj, 158, 753
\bibitem[Ramsay, G. et al. (2004)]{Ramsay04} Ramsay, G., Cropper, M., 
Wu, K., Mason, K.O., Cordova, F.A., Priedhorsky, W., 2004. \mnras, 350, 1373
\bibitem[Reinsch et al.(2003)]{Reinsch03} Reinsch, K., Euchner, F., Beuermann, 
K., Jordan, S., 2003, astro-ph, 0302056
\bibitem[Rousseau, T. et al. (1996)]{Rousseau96} Rousseau, T., Fischer, A., 
Beuermann, K., Woelk, U., 1996, \aap, 310, 526
\bibitem[Schwope et al.(1990)]{Schwope90} Schwope, A.D.  1990, 
Reviews In Modern Astronomy, 3, 44
\bibitem[Schwope et al.(1999)]{Schwope99} Schwope, A.D., Schwarz, R., Greiner, 
J., 1999, \aap, 348, 861
\bibitem[Schwope et al.(2007)]{Schwope07} Schwope, A.D., Staude, A., Koester, 
D., Vogel, J., 2007, \aap, 469, 1027
\bibitem[Szkody, P. et al. (2006)]{Szkody06} Szkody, P., Harrison, T.E., 
Plotkin, R.M., Howell, S.B., Seibert, M., Bianchi, L., 2006, \apj, 646, 147
\bibitem[Tapia, S. (1977a)]{Tapia77a} Tapia, S.  1977a, \apj, 212L, 125
\bibitem[Thompson, A.M.\& Cawthorne, T.V. 1987)]{Thompson87} Thompson, A.M,
Cawthorne, T.V. 1987, \mnras, 224, 425
\bibitem[Thorstensen, J.R. (2003)]{Thorstensen03} Thorstensen, J.R., 
2003, \aj, 126, 3017
\bibitem[Vacca et al.(2003)]{Vacca03} Vacca, W.D., Cushing, M.C., 
Rayner, J.T., 2003, \pasp, 115, 389
\bibitem[Wheatley \& Ramsay(1998)]{Wheatley98} Wheatley, P.J.,
Ramsay, G., 1998, ASP Conference Series, 137, 1998
\bibitem[Wickramasinghe, D.T \& Ferrario, L.(2000)]{Wickramasinghe00} 
Wickramasinghe, D.T , Ferrario, L., 2000, \pasp, 112, 873
\bibitem[Wickramasinghe, D.T. \& Meggitt, S.M.A. (1985)]{Wickramasinghe85a}
Wickramasinghe, D.T., Meggitt, S.M.A., 1985, \mnras, 214, 605
\bibitem[Wickramasinghe, D.T. \& Meggitt, S.M.A. (1985)]{Wickramasinghe85b}
Wickramasinghe, D.T., Meggitt, S.M.A., 1985, \mnras, 216, 857
\end{thebibliography}
\end{document}